\documentclass{sig-alternate-10pt}

\usepackage{times}
\usepackage{amsmath, amssymb}
\usepackage{graphicx}
\usepackage{xspace}
\usepackage{color}    
\usepackage{pifont}
\usepackage{subfigure}
\usepackage[hyphens]{url}
\usepackage{cite}     
\usepackage{algorithm, algorithmic}
\usepackage{comment}
\usepackage{enumitem}
\usepackage{lastpage}
\usepackage{flushend}
\usepackage[font={footnotesize}]{caption} 

\newcommand{\myqed}{\rule{7pt}{7pt}}

\newenvironment{proof-sketch}{\noindent{\bf Sketch of Proof:}\hspace*{1em}}{\myqed\bigskip}
\newenvironment{proof-idea}{\noindent{\bf Proof Idea:}\hspace*{1em}}{\myqed\bigskip}
\newenvironment{proof-of-lemma}[1]{\noindent{\bf Proof of Lemma #1:}\hspace*{1em}}{\myqed\bigskip}
\newenvironment{proof-attempt}{\noindent{\bf Proof Attempt:}\hspace*{1em}}{\myqed\bigskip}

\newcommand{\etal}{{et~al}.\@~}
\newcommand{\eg}{e.g.,\xspace}
\newcommand{\ie}{i.e.,\xspace}

\def\centerhack#1{\hbox to 0pt{\hss\footnotesize #1\hss}}
\def\dchack#1{\vbox to 0pt{\vss{\hbox to 0pt{\hss#1\hss}}\vss}}

\newcounter{linecounter}

\renewcommand{\paragraph}[1]{\vspace{2mm}\noindent{\textbf{#1}}\quad}

\definecolor{orange}{rgb}{1,0.5,0}

\newcommand{\namens}{\textsc{DENA}}
\newcommand{\name}{\textsc{DENA}\xspace}
\newcommand{\names}{\textsc{DENA}s\xspace}
\newcommand{\scion}{\textsc{\textcolor{black}{SCION}}\xspace}
\newcommand{\ete}{\textcolor{black}{end-to-end}\xspace}

\newlist{ReqR}{enumerate}{1}
\setlist[ReqR]{label=\textbf{Req \arabic*.},leftmargin=*,itemindent=1cm}
\makeatletter
\def\@copyrightspace{\relax}
\makeatother

\begin{document}
\sloppy

\title{Bootstrapping Real-world Deployment of\\
Future Internet Architectures}

\author{Taeho Lee$^{\dagger}$ Pawel Szalachowski$^{\dagger}$ David Barrera$^{\dagger}$ Adrian Perrig$^{\dagger}$ Heejo Lee$^{\S}$ David Watrin$^{\ast}$\\
\begin{tabular}{cp{0.1cm}cp{0.1cm}c}
$^{\dagger}$ETH Zurich && $^{\S}$Korea University && $^{\ast}$Swisscom AG\\
\parbox{5cm}{\{kthlee, psz, david.barrera\\ aperrig\}@inf.ethz.ch}
    && heejo@korea.ac.kr && David.Watrin@swisscom.com \\
\end{tabular}
}
\maketitle

\section*{Abstract}
The past decade has seen many proposals for future Internet architectures.
Most of these proposals require substantial changes to the current networking
infrastructure and end-user devices, resulting in a failure to move from theory
to real-world deployment. This paper describes one possible strategy for
bootstrapping the initial deployment of future Internet architectures by
focusing on providing high availability as an incentive for early adopters.
Through large-scale simulation and real-world implementation, we show that with
only a small number of adopting ISPs, customers can obtain high availability
guarantees. We discuss design, implementation, and evaluation of an
availability device that allows customers to bridge into the future Internet
architecture without modifications to their existing infrastructure.

\section{Introduction}
\label{sec:introduction}
To be successful in their deployment, designers of new technologies must not
only solve technical issues (\eg compatibility, efficiency, etc.), but must
also offer convincing incentives for users to invest time and money to adopt
the technology. Without proper incentives, achieving substantial real-world
deployment is likely a futile effort.

The deployment of new Internet architectures is especially challenging given
the millions of hours of effort that have gone into building the networks these
new architectures are designed to replace.  Over the past two decades, a number
of new ideas that provide strong technical advantages over existing networks
and protocols (\eg mobile IP~\cite{rfc5944} and IP multicast~\cite{rfc1112})
have been proposed.  However, they have failed to gain mainstream adoption. The
resistance to change may be justified, since sizeable financial investments and
millions of hours spent on training make embracing potentially disruptive
changes a less than ideal proposition for incumbents. 

Future Internet Architectures (FIAs) aim to solve many problems with
current-generation networks (\eg security, availability, and scalability). To
solve these issues, FIAs suggest using radically different paradigms.  For
example, NDN~\cite{Jacobson:2009:NNC:1658939.1658941} provides efficient data
delivery by treating data (rather than the end-points) as a first-class citizen
on the network; MobilityFirst~\cite{Raychaudhuri:2012:MRT:2412096.2412098}
designs a mobility-centric network; XIA~\cite{Han2012} delivers a flexible
means of evolving the Internet's core; and \scion~\cite{scion2011} and
NIRA~\cite{nira2003} focus on making the Internet more resilient to failures.
Despite the appealing benefits offered by these architectures, they have yet to
gain mainstream adoption.

In this paper we present a case study on how to bootstrap deployment of FIAs by
focusing on how to convince early adopters. We demonstrate that deployment of a
new Internet architecture can provide tangible benefits to early adopters while
requiring minimal changes to existing infrastructure during the initial stage
of deployment.  Our goal is to present a feasible adoption plan to gradually
(and without friction) gain traction. Thus, instead of proposing a generic
one-size-fits-all plan (\ie attempting to convince manufacturers, ISPs,
developers, and end users that they should immediately deploy a particular
FIA), we focus on providing a single tangible benefit to early adopters:
increased network availability.  For this goal, we propose using FIAs as
fallback (backup) networks when specific quality of service metrics are not met
by current Internet paths. We note, however, that we focus exclusively on
initial deployment. Later steps will need other incentives since network
availability alone may be insufficient to motivate wide-scale deployment. 

Based on simulations, real-world implementation, and evaluation, we show that
with only a small number of deploying nodes, new Internet architectures, such
as NIRA and \scion, can transparently improve network availability while
introducing negligible performance overhead once deployed.  Our results provide
evidence that the seemingly impossible task of deploying a new Internet
architecture may, in fact, be possible.

Our contributions are as follows:

\begin{itemize}
\setlength{\itemsep}{-3pt}
	
	\item We describe the value proposition for both end-users and ISPs showing
that our approach benefits from natural scalability and wide-scale customer
base reachability (Section~\ref{sec:background}).

    \item We design and implement \name (Device for ENhancing Availability), a
bump-in-the-wire device that can automatically detect and establish fail-over
data paths over FIAs, and transparently fail-over to those paths when network
quality is low. \name requires no user configuration and is designed to be
minimally intrusive (Sections~\ref{sec:overview} and~\ref{sec:design}).

    \item We demonstrate, through a full-scale BGP simulation over the CAIDA
Inferred AS Relationship dataset (45,942 ASes), that a small number (less than
five) of ISPs deploying a FIA can significantly reduce attacks on availability
(Section~\ref{ssec:simulation}).

\end{itemize}

\section{Related Work}
\label{sec:related}
We review related work in two overarching themes: increasing network
availability and incremental deployment of new architectures. 

Several proposals aim to increase Internet availability. One avenue towards
this objective is through the use of indirection. Andersen \etal\cite{RON2001}
propose Resilient Overlay Network (RON) to shorten recovery time incurred from
BGP outages by using an alternate network path through an overlay. While the
work demonstrates the effectiveness of an overlay against BGP outages, it does
not describe Internet-scale deployment issues where different ISPs can own
different parts of the overlay. Our paper discusses deployment incentives for
possibly competing ISPs. 

Another closely related work is Advertised Reliable Routing Over Waypoints
(ARROW). In their paper, Peter~\etal\cite{DBLP:conf/sigcomm/PeterJZWAK14} show
that by redirecting traffic through tunnels, Internet reliability can be
enhanced. The authors validate their claim through BGP simulations considering
IP prefix hijacking attacks. The work herein corroborates the results of Peter
\etal finding that redirecting traffic to a victim AS using a tunnel provides
higher resilience against prefix hijacking attacks when an adversary targets
the victim's IP prefixes. Our simulation (see Section~6.4) encompasses a more
comprehensive adversary strategy; unlike in ARROW, we analyze the case where
adversaries are capable of attacking the tunnels as well as the victim
AS.\footnote{We have contacted the authors and confirmed the limitations of
adversary model described in Section~5.8 of their paper (\ie the adversary only
announces the prefixes of the victim).}  Additionally, our paper studies
incremental deployment aspects of FIAs, while the ARROW paper considers a
minimal change that can provide the greatest increase in availability. For
instance, the authors do not discuss how and who will manage the
\textit{Internet Atlas}, which is crucial for incremental deployment since such
an atlas likely requires global coordination between the deploying ASes.

There have been proposals to study incremental deployability aspects of an
architecture. In LISP (The Locator Identifier Separation Protocol), Saucez
\etal\cite{LISPDeployment2012} postulate ``there must be a clear deployment
path that benefits early adopters''.  Ratnasamy \etal\cite{Evolvable2005}
distilled the properties that facilitate the deployment of a new architecture.
They propose the use of IP anycast as the means to access the new architecture.
We build upon the insights that Saucez \etal and Ratnasamy \etal have provided
in terms  incremental deployability. However, we avoid using IP anycast as the
medium to access the new architecture because the use of anycast may preclude
ISPs from engaging into a direct business relationship with customers. Instead,
in our proposal an interested customer purchases a value-added service from a
deploying ISP, and the ISP would distribute a gateway device to access its
service.

\section{Background and Motivation}
\label{sec:background}
In this section, we motivate the need for network availability for ISPs and
customers focusing on limitations of the Border Gateway Protocol (BGP). 

\subsection{Demand for Network Availability}

The ever-increasing dependence on the Internet in our society has made
availability a critical requirement for network infrastructure. We define
availability to be the resilience against any type of attack or other network
fault (\eg misconfiguration) that could prevent or decrease the quality of a
network connection.

Due to their core business model (\ie offering connectivity to customers), ISPs
require high network availability. We can infer the existence of this demand by
analyzing the number of multi-homed endpoint ASes on the Internet. Using a
snapshot\footnote{The CAIDA dataset was collected on August 1st, 2014.} of the
CAIDA Inferred AS Relationships dataset~\cite{caida}, we observe that 22,386
out of 38,772 (approximately 57\%\footnote{Note that this ratio could be an
over-estimate as there are single-homed stub ASes without AS numbers.})
endpoint ASes have two or more provider ASes. Of these multi-homed ASes, 819
have five or more provider ASes. While there are reasons other than
availability for ASes to be multi-homed (\eg load-balancing, cost-balancing),
multi-homed ASes are more likely to be accessible even if one of their BGP
paths is unavailable.

Customers span the spectrum from residential end-users to large-scale
enterprises to governments.  A recent study found that residential users are
becoming sensitive to quality of their home Internet
connections~\cite{survey2013}. Businesses,  especially those which are
geographically diverse, make use of Virtual Private Network (VPN) technologies
to interconnect sites. For these businesses, the availability of intranet
resources is tied to the availability of their Internet connectivity.
Governments may want to ensure availability of their critical infrastructure
to control smart grids or to ensure the success of online elections~\cite{hk_poll}. 

\subsection{Degradation of Network Availability}
\label{sec:bgp_resilience}

Several factors may impact network availability on the Internet. These factors
range from transient link failures to route changes to security vulnerabilities
in routing protocols. Attacks on BGP are one of the major contributing factors
to availability issues on the Internet today~\cite{BGPVulnerabilities}.  One of
the most common threats in this context is known as \textit{IP Prefix
Hijacking}. Briefly, a BGP route is defined as an IP prefix and an AS-level
path which leads to that prefix. Thus, IP prefix hijacking denotes the
advertisement (malicious or accidental) of a prefix which is not authorized for
use by an advertising entity and which diverts the BGP path for that prefix to
the hijacker.  Due to the lack of authentication in BGP messages, IP prefix
hijacking is relatively easy to perform as the adversary can announce any
prefix of another AS (the standard threat model for this type of attack). As
a consequence, false injection of update messages can be used to launch
blackhole attacks~\cite{SPV2004} which hijack and drop traffic to compromise
availability.  Although the following does not fundamentally protect BGP
against hijacking attacks, we describe two practices that help in building a
more secure path in BGP.

\begin{itemize}
\setlength{\itemsep}{-3pt}

    \item{\textbf{/24 Prefix Announcements}}: Due to BGP's default route
selection policy, longer prefixes are harder to compromise than shorter
prefixes~\cite{conf/sp/HuM07,conf/sigcomm/BallaniFZ07}.  However, ISPs are
hesitant to accept /24 prefix announcements pervasively because doing so would
dramatically increase the size of routing tables on their border routers.

    \item{\textbf{Shorter BGP Paths}}: Shorter paths are usually more resilient
than longer paths because border routers typically prefer routes with shorter
AS-Path length. As the number of AS hops decreases, there are fewer network
locations (routers) from which an attack could be launched. Thus, if a route
between two end-hosts could be divided into multiple shorter path segments, it
would be more secure than the single longer path between the two end-hosts. We
confirm this effect through BGP simulation in Section~\ref{ssec:simulation}. 

\end{itemize}

\subsection{Value Proposition for Early Adopters}
\label{ssec:approach}
\paragraph{Value for ISPs.} As early adopters, ISPs that deploy an FIA obtain
several benefits.  First, deploying ISPs benefit from higher availability by
increasing their resilience to hijacking attacks (see
Figure~\ref{fig:sim-hijack}); and, the benefit exist even with a small number
(see Figure~\ref{fig:sim-reach}) of deploying ASes.

The benefit is accrued without sacrificing the scalability of BGP routing
tables. If every domain were to announce /24 prefixes for higher availability
(\ie stronger resilience against IP prefix hijacking attacks), the BGP routing
table does not scale.  Instead, in our approach, since only a few initially
deploying ASes announce /24 prefixes to protect their tunnels, the overhead on
BGP routing tables is small. Furthermore, beyond initial deployment, our
approach benefits from natural scalability; as more ISPs deploy the FIA, fewer
/24 announcements are needed since two contiguous ISPs that deploy FIA can
communicate directly without a /24 announcement.  Consequently, we expect the
number of announced /24 tunnel prefixes to increase during the early stage of
deployment and decrease as more ISPs deploy.

Second, there may exist a business incentive for deploying ISPs to offer high
availability to customers of other ISPs (\eg via IP tunnels as described in
Section~\ref{ssec:deployment_model}). Lastly, ISPs can create a niche market
for a set of customers who demand higher availability than BGP can provide but
cannot afford dedicated leased lines.

\paragraph{Value for Customers.} Customers may obtain high-availability (even
if the customer's local ISP does not provide such functionality) by subscribing
to a remote provider via an access tunnel.  More importantly, customers can be
bridged into high-availability architectures without changes to software on
their local devices or modifications to their Internet service. We show one
method to achieve this zero-configuration setup in Section~\ref{sec:overview}. 

\section{Overview and Requirements for a High-availability Device}
\label{sec:overview}
Guaranteeing benefits under partial deployment alone is not sufficient for a
successful initial deployment. Customers must be able to subscribe to an FIA
without carrying out any complicated tasks, such as configuring their network
devices or updating them to a new network stack---updating the network stack on
light bulbs, cameras, televisions, refrigerators, and other Internet of things
devices may be infeasible. To this end, we develop a \textit{bump-in-the-wire}
interface device which we refer to as \name (Device for ENhancing Availability)
to be placed between the customer's network and their Internet provider. An
alternative approach could be to deploy such interface devices at the ISPs
themselves (this approach is used by Peter
\etal~\cite{DBLP:conf/sigcomm/PeterJZWAK14}), completely removing end-user
involvement. However, this strategy would preclude users from subscribing to
the FIA if their immediate ISP does not deploy it.

The primary function of \name is to increase Internet availability to its
subscriber by leveraging a given FIA as a fail-over network. That is, if the IP
path between two customers (deploying end-points) becomes unavailable, \name
would detect the unavailability and automatically fail-over to a path through
the FIA by encapsulating IP packets.

To provide high-availability guarantees, \name devices need to implement four
functionalities. For a given communication flow between a subscriber and a
peer, the \name needs to \textit{1)}~identify the presence of a peer;
\textit{2)}~if present, establish FIA path(s) to be used as fail-over;
\textit{3)}~continuously measure the packet loss rate of the path(s); and
\textit{4)}~fail-over to a path offered by the FIA if necessary. 

Although our design and implementation of \name remains generic for any FIA
that is capable of providing high-availability guarantees (\eg NIRA and
\scion), throughout the following sections, we draw on \scion as the chosen
example FIA. We find it useful to focus on one particular FIA to keep the
discussion concrete and to drive the narrative. In addition, our access to the
global \scion testbed enables for a thorough evaluation of our implementation.

For space reasons, we do not provide details of the \scion architecture in this
paper; however, the discussions in the following sections should be clear
without any prior knowledge of \scion (Zhang \etal~\cite{scion2011}). 

\subsection{ISP Deployment Model}
\label{ssec:deployment_model}

To minimize impact on their current infrastructure, ISPs are likely to deploy a
high-availability infrastructure as an overlay to their current
networks~\cite{peterson2003blueprint}. Customers, as well as other ISPs will
use IP tunnels to bridge into these overlays. 

Hence, constructing reliable tunnels is essential to provide high-availability
guarantees. To this end, /24 prefix blocks are used to reduce the risk of
traffic hijacking attacks against the tunnels. 

\begin{figure}
	\centering
	\includegraphics[width=0.9\linewidth]{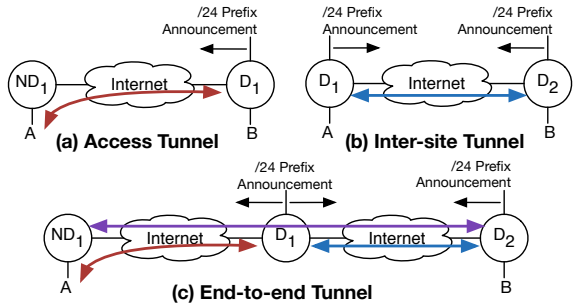}
	
	\caption{Three tunnels that are used in FIA deployment. In each subfigure,
		circles marked D$_i$ represent ASes that have deployed FIA, and circles 
		marked
		ND$_i$ represent non-deploying ASes.}
	
	\label{fig:tunnels}
\end{figure}

For our deployment strategy, we define the following tunnel types (shown in
Figure~\ref{fig:tunnels}), and explain how tunnels are protected against
hijacking by using the practices described in Section~\ref{sec:bgp_resilience}.

\paragraph{Access Tunnel.} A tunnel with which a deploying ISP can offer
high-availability services to customers whose ISPs do not support these
features. As shown in Figure~\ref{fig:tunnels}a, to protect the tunnel between
A and itself from hijacking, the deploying AS (D$_1$) announces a /24 prefix
that  contains the IP address used for that tunnel. This announcement reduces
the probability of an adversary successfully hijacking this tunnel if they are
multiple hops away from A and D$_1$.

\paragraph{Inter-site Tunnel.} A tunnel that connects two non-contiguous
deploying ASes over the Internet (see Figure~\ref{fig:tunnels}). The ASes need
to protect the tunnels that link deploying sites from prefix hijacking attacks.
Similar to the access tunnels above, each of the two ASes announces a /24
prefix block that contains the IP address used as its tunnel end-point address. 

\paragraph{End-to-end Tunnel.} A series of tunnels that consist of access
tunnels and inter-site tunnels to connect customers A and B over the FIA
deployment. For instance, in Figure~\ref{fig:tunnels}, the end-to-end tunnel
consists of one access tunnel and one inter-site tunnel.

\subsection{End-to-End Communication Scenario}
\label{sec:e2e_scenario}

\begin{figure*}
\centering
\includegraphics[width=0.75\textwidth]{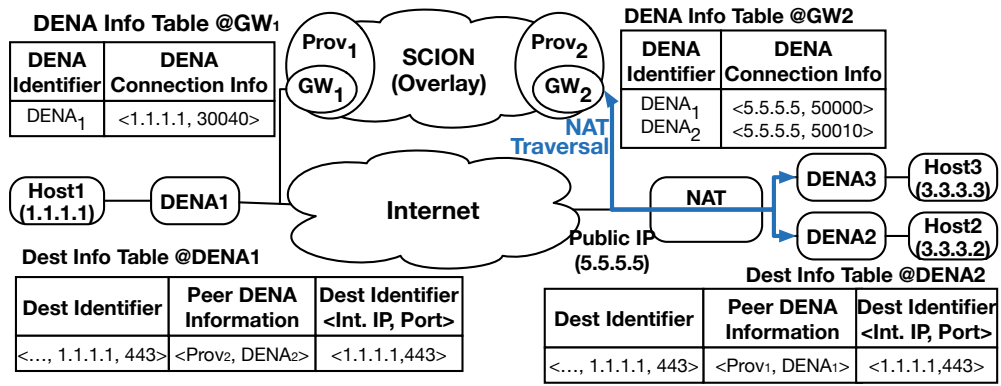}
\caption{An example of end-to-end communication. The destination identifier is
implemented as the 5-tuple information. See Section~\ref{sec:design}.}
\label{fig:e2e_comm}
\end{figure*}

Before describing the design and implementation of \name, we sketch a
high-level picture of how an end-to-end communication flow would behave in a
typical Internet-wide scenario shown in Figure \ref{fig:e2e_comm}.

In this scenario, there are two \scion providers (Prov$_1$, Prov$_2$) that have
deployed gateways (GW$_1$, GW$_2$) to serve their \scion subscribers. There are
three end-hosts (Host$_1$, Host$_2$, Host$_3$), where Host$_1$ has a public IP
address, and Host$_2$ and Host$_3$ are behind a NAT. To enhance their Internet
availability, Host$_1$ has subscribed to the \scion service from Prov$_1$, and
Host$_2$ and Host$_3$ from Prov$_2$. All hosts have placed \name devices as
bump-in-the-wire on their network connection.

Consider a scenario where Host$_2$ initiates communication to Host$_1$. Upon
receiving a packet from Host$_2$, \namens$_2$ checks if it knows the peer \name
associated with the destination address of the packets. If it does not know the
associated peer, \namens$_2$ initiates a Discovery Process
(Section~\ref{subsection:Discovery}) and exchanges bootstrapping information
that is necessary for constructing \scion overlay paths to the remote peer. In
addition, the bootstrapping information contains information necessary for NAT
traversal, which we discuss in Section~\ref{sec:discussion}.

A \name monitors the packet loss  rates of the paths (both IP and \scion paths)
to all of the discovered peer \names
(Section~\ref{subsection:path_management}). Whenever the quality of the IP path
to a peer \name degrades below a given threshold, and the quality of a path
provided by \scion is acceptable, all of the flows associated with the peer
\name are encapsulated and redirected to the \scion path to ensure continued
availability.

\begin{figure}[h]
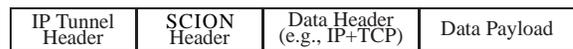

\renewcommand{\arraystretch}{2.5}
\linespread{0.5}\selectfont\centering
\begin{tabular} {|c|c|c|c|}
\hline
\parbox[c][][t]{1.3cm}{\centering{\small IP Tunnel Header}} &
\parbox[c][][t]{1.3cm}{\centering{\small \scion Header}} &
\parbox[c][][t]{1.7cm}{\centering{\small Data Header (\eg IP+TCP)}} &
\parbox[c][][t]{1.7cm}{\centering{\small Data Payload}}\\[1pt]
\hline
\end{tabular}

\caption{Packet structure over \scion}
\label{fig:pkt_structure}
\end{figure}

Redirection over \scion is accomplished via a series of tunnels using the
packet structure shown in Figure \ref{fig:pkt_structure}. The first and last
tunnels are access tunnels (see Figure~\ref{fig:tunnels}a) between the \names
and the \scion gateways.  The intermediate tunnel is a series of inter-site
tunnels (see Figure~\ref{fig:tunnels}b) that links the two \scion ASes to which
the two end-hosts are subscribed. 

In our communication example between Host$_2$ and Host$_1$, the packets from
Host$_2$ to Host$_1$ traverse from \namens$_2$ to \namens$_1$ via tunnels
across \namens$_2$ to GW$_2$, Prov$_2$ to Prov$_1$, and GW$_1$ to \namens$_1$.
\namens$_1$ removes the tunnel header and the SCION header, and forwards the
regular IP packets to Host$_1$.

\subsection{Requirements}
\label{subsection:requirements}

Previous work in the incremental deployment literature~\cite{Evolvable2005,
LISPDeployment2012, Handley2006} has discussed desirable properties for an
architecture to gain adopters. These properties mainly refer to the need for
incentivized adoption, independence from ISPs that hosts buy their Internet
connection from, and compatibility with existing devices and infrastructure.
Based on these properties, we define the design requirements for our own
incremental deployment device.

\begin{ReqR}
\setlength{\itemsep}{-3pt}

    \item{\textbf{No Changes to End-hosts}}: End-hosts should not require any
modification to make use of the FIA. Rather, \name should operate as an
interface between the current Internet and FIA, enabling a smooth transition
between the two architectures.  \label{itm:no-change}

    \item{\textbf{Zero configuration}}: \name should operate as a
``plug-and-play'' device and should not require any configuration by users.
However, similarly to wireless access points, a \name should allow fine-tuning
of parameters by advanced users.  \label{itm:zero-config}

    \item{\textbf{Robustness}}: \name should work under a majority of network
configuration scenarios and in the presence of packet-modifying or
packet-filtering middleboxes (\eg NATs and firewalls).  \label{itm:minimal}

    \item{\textbf{Autonomous peer discovery}}: \name should not require any
additional infrastructure to detect a peer \name. Such additional requirements
would create complications regarding who would build and maintain the
additional infrastructure and these complications will hinder incremental
deployment.  \label{itm:self-disc}

\end{ReqR}

\section{Design and Implementation of DENA}
\label{sec:design}
\begin{figure}[h]
\centering
\includegraphics[width=\columnwidth]{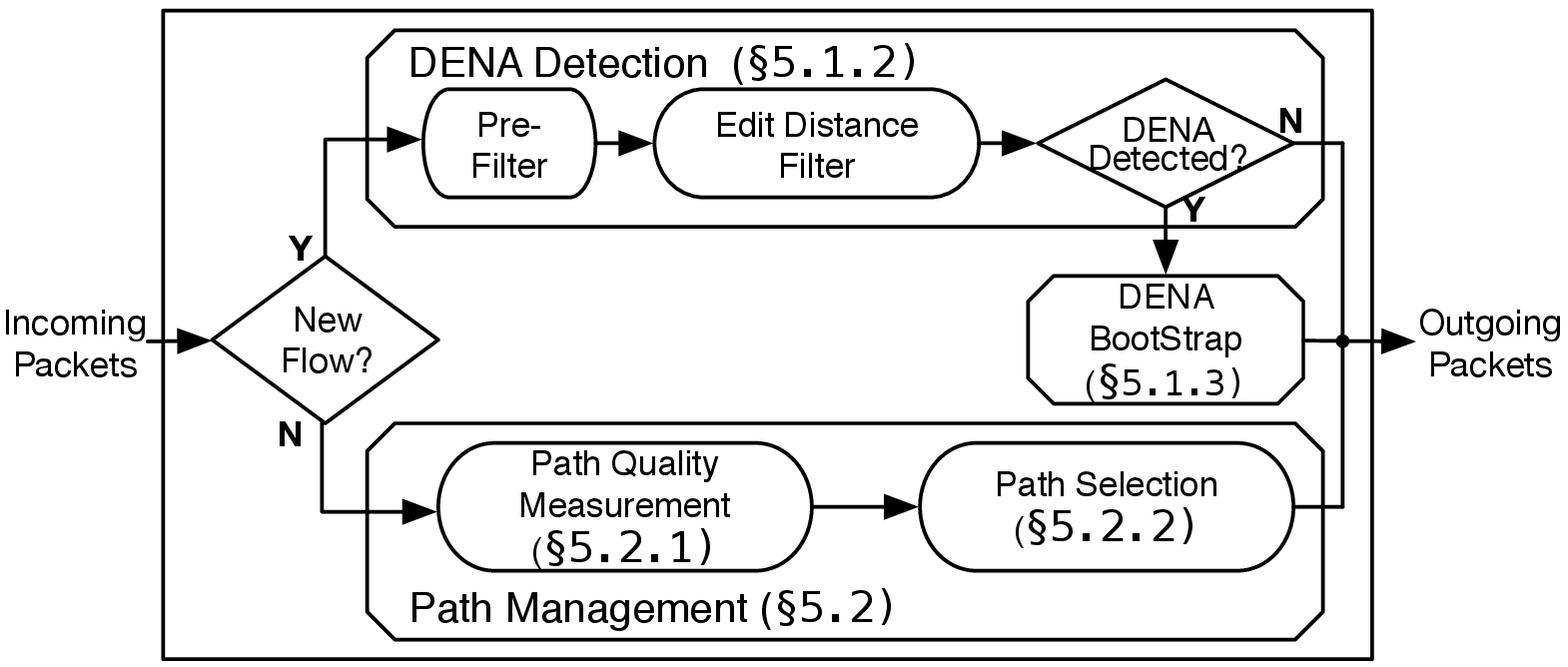}
\caption{Major \name components.}
\label{fig:box_outline}
\end{figure}

This section describes the \name design shown in Figure~\ref{fig:box_outline},
which shows the major components (discovery protocol and path management) and
the corresponding sections where the descriptions for the components can be
found.

\subsection{Discovery Protocol}
\label{subsection:Discovery}

The goals of the \name discovery protocol are \textit{1)}~to automatically and
transparently (\ie without interfering with active data flows) detect the
presence of a remote peer and \textit{2)}~to exchange bootstrapping information
that is necessary for constructing paths over \scion to the peer \name.

Discovery is performed whenever a \name device detects a new communication
flow, identified by the 5-tuple information (source IP, destination IP, source
port, destination port, and protocol number) of the packets in the flow. Ports
are used in addition to the source and destination address pairs so that \names
behind a shared NAT can be distinguished and identified. For example, in
Figure~\ref{fig:e2e_comm}, \namens$_1$ cannot identify whether the
communication flow is being forwarded by \namens$_2$ or \namens$_3$ based only
on the IP address because the packets forwarded by both \names would have the
same source IP address, 5.5.5.5.

Before describing the two parts of the discovery procedure (detection and
bootstrapping), we describe the signaling channel that is fundamental to our
discovery procedure.

\subsubsection{Signaling Channel}
\label{sssec:hidden_signals}

The most important challenge in \name detection is to construct a reliable
signaling channel to the peer \name through which the \name discovery message
is transmitted. Naively, one can define a dedicated UDP or TCP port, and send a
discovery message to that port. Although the approach may work for general
cases, it can be problematic in the presence of middle-boxes such as firewalls
and intrusion detection systems. Additionally, users may be required to perform
manual configurations on such devices, which violates the zero-configuration
requirement from Section~\ref{subsection:requirements}.

Alternatively, we suggest constructing the signaling channel using end-hosts'
traffic. This type of channel has been widely studied in network steganography
with two generic approaches~\cite{Mazurczyk:2013:VSD:2543581.2543587}:
\textit{1)}~manipulating structure of packet streams (\ie temporal variation in
packet transmission times); and \textit{2)}~manipulating packet headers, such
as IP and TCP headers.  Although the signaling channel constructed via the
former method tend to be more robust against various middle-boxes, it
unavoidably introduces undesirable delay and jitter into end-hosts'
communication~\cite{mazurczyk2011}. Thus, we use the latter method to construct
the signaling channel.

To make \name compatible with any IP based protocols, we use the TTL and
Identification (IPID) fields in the IP header. Note that while we are
\textit{hiding} signals inside the IP header, this is done for compatibility
rather than for security. That is, our objective is not to prevent an observer
from seeing that the channel is being established, but rather to prevent the
signaling messages from being parsed (possibly incorrectly) by the destination
device in case a \name device is absent.

\noindent\textbf{TTL.} Although 255 is the literal maximum value of the TTL
field, IP paths on the Internet are generally shorter than 64 router hops (with
64 being the recommended initial TTL value~\cite{rfc1700}). Hence, it is
possible to decode the values that the sender has chosen if the initial TTL
values are chosen to be at least 64 units apart. Since there can only be tuples
of at most three TTL values that are 64 units apart from each other, we define
three signals A, B, C from three TTL values, 64, 128, 192, respectively.

\noindent\textbf{IPID.} Since the use of the IPID field remains unspecified for
unfragmented packets, the field can be used to embed hidden
signals~\cite{rfc6864}. We define three hidden signals from the IPID values so
that the \name detection procedure can be used with both TTL-based and
IPID-based signals.

We choose three sets of IPID values whose 12 most significant bits (MSBs) are
0x001, 0x7FF, 0xFFF as signals A, B, and C, respectively. The values are chosen
to be maximally spaced apart (without including IPID value of zero) to prevent
operating systems that increment the IPID value by one for each transmitted
packet, from accidentally transmitting all three of our defined signals during
the \name detection procedure. In addition, there are 16 IPID values that
correspond to a hidden signal. Although fragmentation is discouraged in
practice (for efficiency reasons), fragmented packets do exist on the
Internet~\cite{frag2001}. The last four bits can change to repeat our signals
even when some of the IPID values are used for fragmented packets.

\subsubsection{DENA Detection}
\label{subsubsection:detection}

For detection, every \name device periodically announces the discovery messages
and responds to the received discovery messages with the bootstrapping
information (defined in Section~\ref{subsubsection:bootstrapping}).

The discovery message is defined as a sequence of signals known to all \names
\textit{a priori} (See Table~\ref{tbl:settings} in
Section~\ref{sec:evaluation}). Because the message is a sequence of hidden
signals, the message is encoded into a stream of consecutive packets. Hence,
when announcing the discovery message, a \name modifies the TTL and the IP ID
fields of the packets that it is forwarding. In our design, both TTL and IPID
fields are used to encode the same hidden signals to enhance the robustness of
\name detection. Thus, even when one of the fields gets scrubbed by
middle-boxes, \names can use the other signal to decode the discovery
message.\footnote{During our experiment, we encountered a middle-box that
scrubs the TTL fields but keeps the IPID fields intact for TCP flows.}

In addition to periodically emitting the discovery message, \name listens for
the discovery message by examining the TTL and IPID fields from streams of
incoming packets. Identifying the discovery message from a stream of packets is
a challenge because the Internet, as a communication channel, may reorder, lose
or duplicate packets. In fact, the decoding problem at hand is similar to the
well-known insertion-deletion channel problem to which an efficient decoding
algorithm is not yet found~\cite{deletion2008}.

Combinatorial methods, which compute the difference between the transmitted and
received messages, have been studied for decoding under insertion-deletion
channel. Edit distance (\ie Levenhstein distance~\cite{Levenshtein}) provides a
metric to compare the difference between the two messages based on the minimal
number of edits (insertion, deletion, replacement) that are necessary to
transform one message to the other. In our design, \name concludes the presence
of a peer if the edit distance between the discovery message and the received
message is below the detection threshold value (\ie fewer number of edits than
the detection threshold).

A \name may falsely conclude the presence of a peer \name. On one hand, a \name
may fail to detect a remote peer that is present (\ie false negative) because
the Internet may drop or reorder the discovery message that the \name sends to
its peer. On the other hand, a \name may falsely detect a peer that is not
present (\ie false positive) due to the IPID fields in a stream of packets
accidentally having the right sequence as the discovery message. There are
trade-offs between these two probabilities, and they depend on the decoding
parameters, such as the length of the discovery message as well as the
detection threshold value.

When the decoding is solely based on edit distance, the false positive
probability can be high. To lower the false positives, we place a light-weight
pre-filter that checks the approximate structure of the received message before
computing the edit distance. The pre-filter lowers the false positive
probability while increasing the probabilities of the false negatives. For an
example design of the pre-filter and the detection error analysis, see
Section~\ref{sec:evaluation}.

\subsubsection{DENA Bootstrapping}
\label{subsubsection:bootstrapping}

After detecting the presence of a peer, bootstrapping information that is
necessary for constructing \scion paths is transmitted. The information can be
transmitted through the signaling channel described in
Section~\ref{sssec:hidden_signals}. However, since the signaling channel has
very small capacity (\ie log$_2$(3) bits per packet\footnote{Assuming perfect
channel that does not lose, reorder, or duplicate packets.}), the channel's
capacity is too small to send the bootstrapping information. (\ie 10 bytes are
needed for \scion).

Alternatively, we define an out-of-band \textit{control-packet} that is
generated by the \names to communicate bootstrapping information. Since a \name
has detected the presence of a peer, it can send a dedicated packet knowing
that it will be parsed and discarded by the peer \name before the packet
reaches the peer end-host. Hence, the control packets do not disrupt end-host
communication.

The control packet must be carefully constructed so that it can be reliably
delivered to the intended \name, passing through middle-boxes that may be
present. To this end, control packets are created by duplicating the end-hosts'
packets. For UDP packets, the fact that end-hosts' packets can reach the
destination ensures that the control packet would be able to reach the peer
\name. For TCP packets, middle-boxes have no choice but to accept the control
packets, as it cannot tell whether the control packets are generated due to
routine packet re-transmissions in the TCP protocol.

After duplicating end-host's packet, the payload (after the IP and transport
header) is modified as follows. A \name identifier message is added at the
beginning to indicate that the message is a \name control packet. Then the
message type to indicate the type of the control message and the message
payload are added.

\subsection{Path Management}
\label{subsection:path_management}

To achieve a high level of availability, it is necessary to periodically
monitor the status and quality of both the IP path and the \scion paths.
Whenever the quality of the IP path degrades below a threshold, communication
between end-hosts is switched over to a higher-quality path offered by \scion.
Furthermore, even after the switch, the IP path is constantly monitored so that
the communication may switch back to the IP path after it is again reliable.

\subsubsection{Path Quality Measurement}
\label{subsubsection:path_measurement}

A \name periodically measures packet loss rates for both IP and \scion paths to
all peer \names found during the \name Discovery Procedure. Unlike the \name
Discovery Procedure, which is performed per flow, path measurement is done per
peer \name.  An advantage of this approach is that it enables more accurate
measurement, since a \name can have many associated flows (\eg multiple
concurrent TCP connections to an HTTP server).

\paragraph{Measurement of the Active Path.} For path quality measurement of the
active path, we rely on data packets from end-hosts that are flowing between a
pair of \names. We adopt a packet loss measurement method that has been used
for quality measurements in MPLS networks~\cite{rfc6374}. Briefly, the proposal
uses a pair of measuring points each of which maintains two counters for
counting incoming and outgoing packets, and computes packet loss rates based on
the four counters.

Specifically, the measurement proceeds as follows between two measuring points,
\textit{A} and \textit{B}. First, \textit{A} initiates the measurement by
sending a request message to \textit{B} with the outgoing packet counter value
at the moment that the request is made. Then \textit{B} replies to the request
with the value of the incoming packet counter as well as its outgoing packet
counter value. Upon receiving the reply, \textit{A} marks its value of the
incoming packet counter. Finally, using the four counter values, \textit{A}
computes the packet loss rate, which is defined to be the maximum of the loss
rates between incoming and outgoing paths.

When computing the packet loss rate, the difference in counter values from
current and past measurements are used. Thus, the two measurement points
\textit{A} and \textit{B} do not need to synchronize the starting time at which
both points start counting outgoing and incoming packets.

For our context, the two measurement points would be the pair of \names. In
addition, the measurement request and reply packets needed for the measurement
are implemented using the \name control messages described in
Section~\ref{subsubsection:detection}.

The MPLS measurement approach offers several advantages for our implementation.
First, it is light-weight\footnote{The measurement points keep trivial amount
of states---the four counter values for the previous measurement.}; second,
synchronization between two \names is not necessary; third, two-way channel
delay can be computed with only a small additional overhead (see Frost
\etal~\cite{rfc6374} for details).

\paragraph{Measurement for Fail-over Paths.} The quality of the fail-over paths
must be monitored so that the \name can ensure that the fail-over paths are
healthy when attempting to send data traffic over one of these paths.  However,
the passive MPLS measurement cannot be used directly for the fail-over path
since no data traverse them.

To overcome such problem, we sample the traffic flowing though the active path
and transmit this traffic over the fail-over paths. Since the loss rate of both
outgoing and incoming traffic are necessary, we need bi-directional traffic
flowing through the fail-over paths. Hence, the two \names need to coordinate
the time when data traffic through the active path is replicated on the
fail-over paths.

To coordinate the measurement intervals between two \names, we use the \name
control channel (Section~\ref{subsubsection:bootstrapping}). Two \names
exchange a ``start message'' to start the measurement and a ``stop message'' to
stop the measurement. An advantage of this approach is that it requires no form
of time synchronization, which can be difficult to achieve without introducing
additional complexity.

\begin{figure}[!htb]
\centering
\includegraphics[width=\columnwidth]{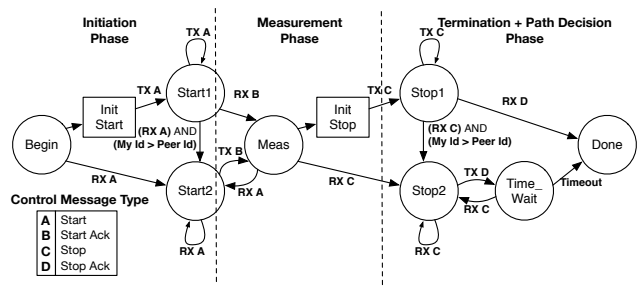}
\caption{State machine for the Path Quality Measurement.}
\label{fig:native_measurement}
\end{figure}

Figure \ref{fig:native_measurement} shows the state machine for the path
quality measurement, which consists of three phases. The Initiation Phase is
started when \name transmits the \textbf{Start} message and is completed when
the initiating device receives a \textbf{Start ACK} message from a peer.  The
initiating and responding \names are decided based on the ID values that they
exchange during the bootstrapping phase.

During the Measurement Phase, each \name performs the measurement to compute
the packet loss rates of the fail-over paths.  Once the measurement is
complete, the two peers listen for \textbf{Stop} messages that are used for
terminating the measurement. In addition, both \names maintain timers to mark
the end of the measurement period. A \name sends a \textbf{Stop} message to its
peer when the timer expires. Upon successfully exchanging the \textbf{Stop} and
\textbf{Stop ACK} messages between the two \names, the path measurement cycle
is completed.

In addition to marking the end of an measurement interval, the \textbf{Stop}
and \textbf{Stop ACK} messages serve another purpose. These messages contain
the packet loss rates and their decisions regarding which path to use to
forward end-hosts' data traffic.

Lastly, similar to the TCP connection tear-down procedure, we implement a
timeout state (\ie \textit{Time\_Wait} state) so that a \name that sends the
\textbf{Stop ACK} is sure that the peer \name has received the \textbf{Stop
ACK} message.

\paragraph{Path Keep-Alive based on Active Probing} In cases where a path drops
all packets (\eg link failure), our measurement does not work. To cope with
such failures, a \name periodically exchanges ICMP probe messages with the peer
\name through all available paths. If a \name does not receive any response for
consecutive probe messages from a path, the \name considers the path to be
unavailable. Additionally, the \name immediately terminates the ongoing path
measurement and starts a new measurement.

\subsubsection {Path Selection}

Path selection is performed whenever a path measurement is completed (when
\name reaches the \textit{Done} state). A new path is selected as follows:
\textit{1)} whenever the loss rate of the IP path is lower than the threshold,
a \name chooses the IP path, \textit{2)} if the loss rate is higher than the
threshold and if there is a \scion path that has lower packet loss rate than
the threshold the \scion path is chosen, \textit{3)} if none of the paths has
lower packet loss rate than the threshold, the path with the lowest loss rate
is chosen.

\subsection{Implementation}
\label{sec:implementation}

The proof-of-concept prototype of the \name consists of 3K lines of C code. The
implementation runs on Linux, using the Netfilter Queue (\texttt{nfqueue})
Library~\cite{nfqueue} which enables IP tables firewall decisions to be made in
user-space. We use the \texttt{nfqueue} library, which allows packet
modification before accepting a packet, to manipulate packets and encode the
hidden signals. \name control messages also make use of the library. 

\section{Evaluation of DENA Design}
\label{sec:evaluation}
We validate our design and implementation of \name through simulation and
real-world evaluation over the \scion testbed.\footnote{The topology of the
testbed can be found at
\url{https://polybox.ethz.ch/public.php?service=files&t=1d1c25c87b2a414dade07c717eb5647c}.}
More specifically, we are interested in \textit{1)}~ensuring our implementation
meets design objectives; \textit{2)}~confirming that the FIA measuring the
false positive and false negative probabilities during \name detection; and
\textit{3)}~evaluating the effectiveness of the path quality measurement
described in Section~\ref{subsubsection:path_measurement}. 

\subsection{Conformance to Requirements}

\names are deployed between the customer's end-hosts and their Internet
connection, which requires no changes to the end-host itself (Req 1). \names do
not require any manual configuration (such as IP addresses) by users. Once
configured by ISPs, \names are designed to be plug-and-play (Req 2). 

Although \names generate control packets, these packets by design are not
forwarded to end-hosts. Thus, end-hosts should never receive or need to parse
control packets. \names can work in the presence of various types of
middle-boxes and NATs, which satisfies the robustness requirement (Req 3).
Furthermore, as control packets (which are opportunistically duplicated from
end-host traffic) are indistinguishable from data packets at the IP and
transport layers, they can be forwarded by middle-boxes without being filtered.

Finally, the \name Discovery Procedure satisfies the self-discovery requirement
(Req 4), since it does not require additional infrastructure for peer \name
detection.

\subsection{DENA Detection Procedure Analysis}
\label{sec:fn_fp_analysis}

As discussed in Section~\ref{subsubsection:detection}, the proposed \name
detection procedure is subject to both false positives (\ie detecting a \name
when it is not present) and false negatives (\ie not detecting a \name when it
is present). In this section we evaluate false positive and false negative
probabilities through simulation. We summarize the parameters that we use for
the \name simulation and implementation in Table \ref{tbl:settings}.

\begin{table}[h]
    \renewcommand{\arraystretch}{1.2}
    \footnotesize
\begin{tabular}{|l|l|}
\hline
\parbox{2cm}{\vspace{3pt}Discovery\\ Message}
& \parbox{5.5cm}{\textbf{\footnotesize ABABABABCCCCCCCCABABABAB}} \\ \hline
\parbox{2cm}{Pre-Filter}
& \parbox{5.5cm}{\vspace{3pt}\textbf{1.} Divide decoded message into 3 blocks of 8 signals.\\
\textbf{2.} Check if there are at least three \textit{A}-signals and three \textit{B}-signals each in the first and the third blocks\\
\textbf{3.} Check if there are at least six \textit{C}-signals in the second
block\vspace{2pt}} \\ \hline
\parbox{2cm}{\vspace{3pt}Edit distance\\ Threshold \vspace{3pt}}
	& 3 or 5 \\ \hline
Meas. Interval & 1 seconds \\ \hline
Meas. Period & 1 seconds \\ \hline
\parbox{2cm}{\vspace{3pt}Path Switch\\ Thresh.\vspace{3pt}}
	& 5\% Packet Loss\\
\hline
\end{tabular}
\caption{Parameter settings for the \name simulation and implementation.}
\label{tbl:settings}
\end{table}

\begin{figure}[h]
\centering
\includegraphics[width=0.9\columnwidth]{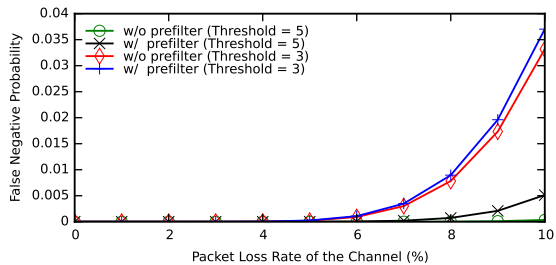}
\caption{Probability that \name fails to detect it a remote peer within five
attempts of the detection procedure.}
\label{fig:false_negative}
\end{figure}

Figure \ref{fig:false_negative} shows the probability that a \name fails to
detect the presence of a peer as packet loss rates are varied at 1\% increments
from 0\% to 10\% within five attempts of \name detection. For precision, we
take the average of one million runs for each fixed packet loss rate. As
expected, the false negative probability increases with increased packet loss
rate. In addition, the false negative probability is higher when the pre-filter
is used or when the lower edit distance threshold value is used.

Furthermore, false negative probability decreases exponentially as the \name
Detection is performed multiple times (not shown in Figure). For example, at
10\% packet loss rate, there is about 35\% chance that a \name fails to detect
a remote peer if the detection procedure is performed only once for the case
with pre-filter and the detection threshold of three.  However, when the
detection procedure is repeated for five times, the false negative probability
decreases to about 0.6\% as shown in Figure~\ref{fig:false_negative}.

A trade-off exists between missed detection and incorrect detection. That is,
the settings that have yielded higher false negative probability have lower
false positive detection probability. To simulate this effect, we input a
stream of packets that consists of one million packets where a random IPID
value is chosen for each of the one million packets. Then, we check whether a
\name would incorrectly identify the presence of a peer within the stream. We
repeat this simulation 10,000 times to get an estimate of the false positive
probability, and the results are summarized in Table \ref{fig:false_positive}.

The false positive probabilities are lower when a pre-filter is used. Also, the
false positive probabilities are lower for the lower threshold value as the
received message has to be closer (\ie fewer number of edits) to the discovery
message. Note that during regular operation, \name would not attempt more than
a few detection attempts. Thus, the values in Table~\ref{fig:false_positive}
represent a worst-case scenario. For our next analysis, we use the threshold
value of 5 with the prefilter as it strikes the balance between FP and FN
probabilities.

\begin{table}[h]
\centering
\footnotesize
\begin{tabular}{|c|c|c|}
\hline
\textit{Threshold} & w/ prefilter & w/o prefilter \\
\hline
3 & 0.02\% & 0.03\% \\
5 & 4.4\% & 11.8\% \\
\hline
\end{tabular}
\caption{False positive probability of the \name Detection Procedure against a stream of one million packets with random IPID values.}
\label{fig:false_positive}
\end{table}

\subsection{Path Switching Based on Path Quality Measurement}

The purpose of this experiment is to confirm that our \name implementation can
react to the change in path quality and switch the end-hosts' traffic to
alternate paths if necessary. For the experiment setup, we deployed two \names
on the \scion global testbed. One was placed at ETH Z\"urich (AS6) and the
other at CMU (AS14). We placed a WANem Emulator~\cite{wanem} on the IP path
from AS6 to AS14, which allowed us to introduce arbitrary packet loss rates
into the IP path.

We deploy an end-host that runs an HTTP server in AS14, and deploy an end-host
that downloads (via the wget application) a large file from the HTTP server in
AS6. At approximately 80 seconds into the file download, we introduce 10\%
packet loss on the IP path, and then remove the packet loss at around 110
seconds.

\begin{figure}
\centering
\includegraphics[width=\columnwidth]{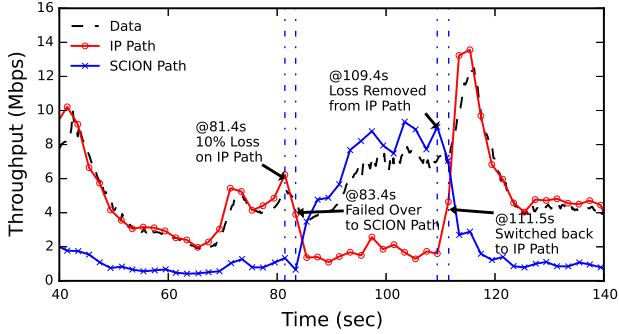}
\caption{Path switching based on measured path quality.}
\label{fig:recovery}
\end{figure}

At the end-host in AS6, we record the throughput of the file download as well
as the data rates on both paths to validate our implementation. Throughput
information is plotted in Figure \ref{fig:recovery}, showing the change in
throughput over time as for the wget application as well as the two paths.

Initially, the end-host's traffic flows through the IP path. In addition, there
is about 10\% of the traffic flowing through the \scion path for measuring the
path loss rate of the fail-over path. When 10\% packet loss is introduced on
the IP path at around 81 seconds, the two \names switch over to the \scion path
about 2 seconds into the degradation. The switch can be verified by the
increase in throughput at the \scion path to that of the actual file-transfer
rate as well as the decrease in throughput at the IP path.

Finally, after the 10\% packet loss is removed from the IP path, the \names
move the end-hosts' traffic back to the IP path after about two seconds of
delay. This test was repeated 5 times obtaining comparable results in each
iteration.

\section{Deployment Analysis}
\label{ssec:simulation}
In the previous section, we have shown that \name can improve the availability
of the end-hosts' connections despite changes in network conditions (\eg when
there is a packet loss in the Internet). In this section, we evaluate the
availability benefits of a FIA when there are adversaries that perform IP
prefix hijacking attacks. In addition, we analyze the benefits that deploying
ISPs could accrue.

To evaluate the deployability and availability benefits of a FIA, we perform
series of BGP simulations by extending the BSIM simulator~\cite{PG-BGP}, where
the BGP paths are computed using route selection based on the standard BGP
routing policies (Gao-Rexford Model~\cite{gao_rexford}). As our topology
dataset, we use a recent snapshot of the CAIDA Inferred AS Relationship
dataset.

\subsection{Tunneled Paths Resilience}
\label{sssec:tun_resiliency}

The announcement of /24 prefixes for the tunnels offers protection of the
tunnels against prefix hijacking attacks; however, the /24 prefix announcements
cannot prevent all hijacking attacks. As the length (in AS hops) of the tunnel
increases, the resiliency against hijacking decreases.  In this section, we
investigate the benefit (\ie resilience against prefix hijacking attack) that a
path constructed using a series of short tunnels can provide over a single BGP
path.

\begin{figure*}[!ht]
	\center
	\includegraphics[width=0.95\linewidth]{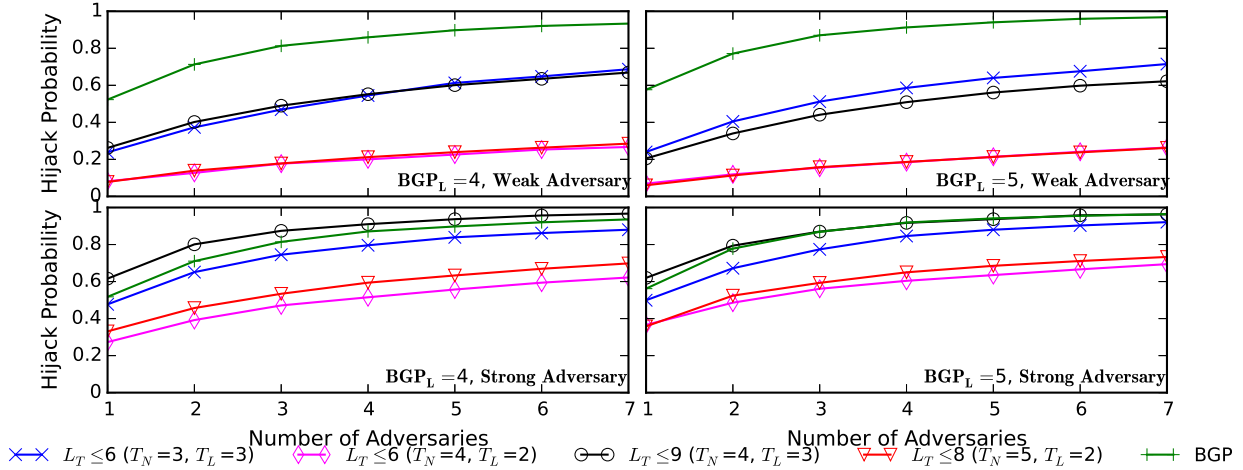}

	\caption{Probability of hijacking BGP and FIA paths under four tunnel
settings while varying the number of adversaries. The upper and lower halves
show the results for weak and strong adversary models respectively; and the
left and right halves show the results for $L_{BGP}=4$ and $L_{BGP}=5$, respectively.}

	\label{fig:sim-hijack}                                                      
	 
\end{figure*}

For our study, we use the following notation:

\begin{description}
        \footnotesize
	\setlength{\itemsep}{-3pt}

	\item[$(AS_x, AS_y)$:] BGP path (list of ASes) between $AS_x$ and $AS_y$,

	\item[$|(AS_x, AS_y)|$:] length (expressed in AS-level links) of $(AS_x,
	AS_y)$ path,

	\item[$T_N$:] number of deploying ASes $AS_{1}, AS_{2}, ..., AS_{T_N}$ 
	that form the overlay \ete tunnel,

    \item[$T_L$:] length of the longest tunnel segment (\ie maximum $|(AS_{i},
        AS_{i+1})|\quad$ for $i\in[1,T_N-1]$),

	\item[$L_{BGP}$:] length of the BGP path between $AS_1$ and $AS_{T_N}$ \\(\ie $|(AS_1,AS_{T_N})|$),

    \item[$L_{T}$:] length of the tunneled path betweeen $AS_1$ and $AS_{T_N}$
        \\(\ie $\sum\limits_{i=1}^{T_N-1}|(AS_i,AS_{i+1})|$).

\end{description}

We assume that the first node of an \ete tunnel path ($AS_1$) is the source
while the last ($AS_{T_N}$) is the destination.  Then, $L_{BGP}$ expresses the
length of the BGP path between source and destination.  We also assume that
traffic from source to destination over the \ete tunnel traverses overlay nodes
$AS_2, AS_3, ..., AS_{T_N-1}$ in that order. The tunnel's path on an AS-level
is a concatenation of BGP paths: $(AS_{1}, AS_{2}),(AS_{2},
AS_{3})...,(AS_{T_N-1}, AS_{T_N})$.

For our simulation, we consider two adversary strategies designed to hijack
traffic from source to destination: \textit{1)} weak adversary which announces
only the destination's prefix, and \textit{2)} strong adversary which announces
all prefixes of $AS_{2}, AS_{3}, ..., AS_{T_N}$.  In both cases an adversary
launches attacks from a randomly compromised AS, however he cannot compromise
ASes on the path between source and destination. Note that
ARROW~\cite{DBLP:conf/sigcomm/PeterJZWAK14} only considers the weak adversary
model and does not consider the strong adversary model.

In this experiment, we simulate 8 scenarios by varying $T_N$, $T_L$, and
$L_{BGP}$ to analyze the resilience that tunnels can provide against prefix
hijacking attacks. For each scenario, while incrementing the number of
adversarial ASes from 1 to 7, we construct and simulate 1000 random and unique
tunnel deployments, where $T_1$ and $T_N$ are randomly chosen from multi-homed
stub ASes and the other tunnel nodes are chosen from all other ASes.\footnote{
To build more confidence of our results, we have run four independent sets of
1000 deployments for all 8 scenarios and confirmed that the results were
similar across the four sets.}

Figure~\ref{fig:sim-hijack} summarizes the results of our simulation. In each
graph in the figure, the $x$-axis represents the number of adversaries (varied
from 1 to 7) and the $y$-axis represents probability values that an attack on
the tunneled path will be successful. The two figures on the left show the
hijack probability of the tunneled path for source and destination AS pairs
that are four BGP hops apart ($L_{BGP}$=4); and, on the right five BGP hops
apart ($L_{BGP}$=5). Moreover, the upper two figures show the results against
weak adversaries while the lower two figures show the results against strong
adversaries. Lastly, each plot also shows hijack probability for the BGP paths
(green line with plus markers).

Our results show that the hijack probability increases as the number of
adversaries increases and that the probability is higher for the strong
adversary model than the weak adversary model. Furthermore, against the weak
adversary model, the probabilities of hijacking the tunneled paths are similar
for the two cases that have the same tunnel segment length (\ie $T_L$) but
different total length (\ie $L_T$)---on the upper two figures, the purple lines
with diamond makers and the red line with inverted triangle markers almost
overlap with each other. This is because the weak adversary model can only
attack the last tunnel segment (\ie ($T_{N-1}, T_N$) as the weak adversaries
only announce the prefix of the destination (\ie $T_N$).

The results show that the tunneled paths have lower hijack probability than the
BGP paths even if the length of the tunneled path (\ie $L_T$) is longer than
that of the BGP paths (\ie $L_{BGP}$). But, the result also shows that if the
length of the tunneled paths becomes too long (\eg twice the length of the BGP
paths), the tunneled paths become more susceptible to hijacking attacks.
However, in practice, it is highly unlikely that the tunneled paths would be
twice the length of the BGP paths.

Lastly, the results show that the composition of the tunneled path affects the
resilience against prefix hijacking attacks: a tunneled path that is composed
of shorter individual segments is more resilient than a path that is composed
of longer individual segments. For the two cases where $L_T\leq6$, the hijack
probability is significantly lower when the length of the individual tunnel is
kept shorter. In Figure~\ref{fig:sim-hijack}, this can be seen by comparing the
blue line with 'x' markers and the purple line with diamond markers.  Moreover,
the result shows that the tunneled paths that have longer total length but
shorter individual tunnel segments (\ie $L_T\leq8$, red lines with inverted
triangle markers) is more resilient than the tunneled paths that have shorter
total length but longer individual segments (\ie $L_T\leq6$, blue lines with
'x' markers).

\subsection{Potential Customer Base}
\label{sssec:coverage}

In this section, we investigate the proportion of multi-homed stub ASes that
are \textit{N}-hops (\ie $T_L=N$) away from any of the deploying ASes as we
vary the number of ASes which deploy the FIA. This experiment should assist in
quantifying the potential customer base (measured in number of ASes) as the FIA
deployment increases.

\begin{figure}[h!]
	\center
	\includegraphics[width=1.\columnwidth]{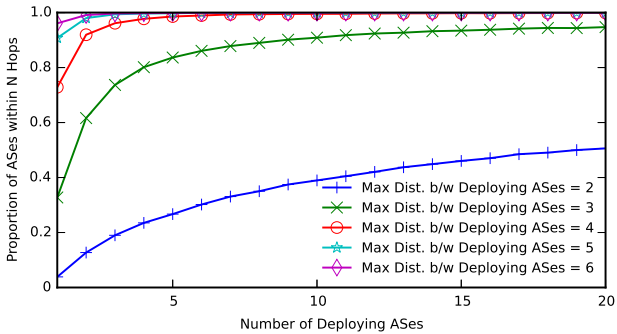}
	\caption{Proportion of ASes that can communicate over FIA.}
	\label{fig:sim-reach}                                                       
	 
\end{figure}

In this experiment, we randomly choose ASes to deploy the FIA such that the
distance between deploying ASes would be at most \textit{N} BGP hops
(\ie$T_L=N$).  Then, for all multi-homed stub ASes, we evaluate the number of
ASes that are within \textit{N} hops from any of the FIA deploying ASes. The
two steps are repeated 5,000 times, and we report the average in Figure
~\ref{fig:sim-reach}, which shows the proportion of the ASes that could be
reached within N hops from the FIA deployment as number of deploying AS changes
from 1 to 20.

Our results show that for higher values of $T_L$, a larger portion of the
multi-homed stub ASes were within the reach of the deploying ASes. Combining
the results from Section~\ref{sssec:tun_resiliency}, we can conclude that for
large values of $T_L$, deploying ASes obtain larger customer-base at the
expense of availability guarantees against prefix hijacking attacks.
Conversely, when offering higher availability guarantees, the customer-base
becomes much smaller (\ie $T_L=2$).

As expected, the number of deploying ASes increases with the size of customer
base. In fact, when there are 20 deploying ASes, the customer-base nearly spans
across the entire Internet for the cases with $T_L\geq4$. In addition, even
with only one deploying AS, there are multi-homed stub ASes that could benefit
from the FIA deployment, and for the cases with $T_L\geq4$, more than half of
the multi-homed stub ASes could access the FIA deployment within $T_L$ hops.

Lastly, Figure~\ref{fig:sim-reach} offers another interesting
observation---reduced benefit (in terms of increase in customer base) as more
ASes participate in FIA deployment. This can be inferred from the gradient of
the plots, which tends to zero as \textit{N} increases. Hence, the result
suggests that beyond the initial phase of the deployment, a different incentive
model may be necessary to further drive the FIA deployment.

\section{Discussion}
\label{sec:discussion}
\subsection{Practical Implications on Deployment}

\paragraph{NAT Implications.} As discussed in
Section~\ref{subsection:requirements}, a \name should work in the presence of
middle-boxes such as NATs. When a \name is placed in-between a NAT and an
end-host (see \namens$_2$ in Figure \ref{fig:e2e_comm}), communication over the
\scion path may not work. For instance, when the communication between Host$_1$
and Host$_2$ in Figure \ref{fig:e2e_comm} is carried out through the \scion
path, \namens$_2$ cannot deliver Host$_1$'s packets to Host$_2$ without being
given Host$_2$'s internal address.

To address this problem, when \namens$_2$ and \namens$_1$ exchange bootstrapping
information during the \name Discovery Process, \namens$_2$ includes the private
address of Host$_2$.

Another problem is the communication between GW$_2$ and \namens$_2$. Because of
NAT, GW$_2$ cannot send a packet to \namens$_2$ before \namens$_2$ starts
forwarding packets via \scion. To avoid this problem, \namens$_2$ periodically
exchanges keep-alive messages to its assigned gateway, performing NAT
hole-punching. Since the \name itself does not have an IP address, it borrows
one of the IP addresses it has learned from the end-hosts to perform the
hole-punching.

\paragraph{MTU Implications.} The measurement and correct configuration of MTU
size is critical for any scheme that requires encapsulation, such as the
tunnels used in our system. When end-hosts perform the path MTU discovery
protocol (PMTUD~\cite{rfc1191}), \name can respond with an MTU size small
enough to provide enough space for encapsulation. In the event that an end-host
does not respond to PMTUD messages (fewer than 10\% of devices do not
~\cite{Luckie2010}), packet fragmentation would be required, increasing the
complexity.

\subsection{Device Configuration and Distribution}

Although \name does not require any configuration by end-users, it requires
configuration by ISPs prior to distribution to customers.  Configuration would
include information necessary for bootstrapping the \name into the FIA
deployment (\ie in case of \scion, this includes \scion ISD ID, AS
AID\footnote{In SCION, each AS is identified by the ID of the ISolation Domain
(ISD) that the AS is in as well as the identifier of the AS (AID).}, and IP
address of the gateway that a \name would contact to access the \scion
network). We note that such configurations are common in today's Internet when
ISPs distribute cable or DSL modems to their subscribers. In addition, for
customers that already have access devices (\ie cable or DSL modems), the
firmwares of these devices could be updated to support \name functionality.

\subsection{Applicability to Other FIAs}

The \name is a generic device that creates tunnels to interconnect two FIA
deploying islands. Hence, it is possible to use \name for other FIAs, as long
as tunneling can be used to interconnect two FIA deploying islands. The
important part in bootstrapping a FIA is to identify the proper incentives for
the early adopters. For instance, for more efficient distribution and access to
the content to early adopters, NDN could be deployed. In NDN deployment, \name
would translate user data requests into an interest packet and send it to the
NDN deployment. When the requested data arrives at the \name, it delivers the
data to the user.

\section{Conclusion}
\label{sec:conclusion}

We have motivated, designed, and evaluated an incremental deployment strategy
for Future Internet Architectures. Through both simulation and real-world
implementation and testing, we demonstrated tangible availability improvements
for deploying ISPs with comparatively little deployment cost. Our evaluation
used \scion to assess the feasibility of our proposal, but the proposed
incremental deployment strategy remains compatible with other FIAs. 

While availability alone will likely be insufficient to motivate full
wide-scale deployment, this paper focuses on convincing early adopters to use a
new architecture. Once a base set of ISPs and customers have deployed a FIA, a
new strategy may be necessary to encourage a majority of ISPs to deploy. For
example, the next set of deploying users could be interested in mobility or
content-centric networking, but such analysis is left open to future research.

\section{Acknowledgements}

We thank Qi Li and Yao Shen for their assistance during the early stage of the
project. We also thank Stephanos Matsumoto for interesting discussions and
feedback on the project. The research leading to these results has received
funding from Swisscom, and from the European Research Council under the
European Union's Seventh Framework Programme (FP7/2007-2013) / ERC grant
agreement 617605. We gratefully acknowledge their support.

\bibliographystyle{abbrv}
\bibliography{0-string,paper}

\end{document}